\input{epsf}

\documentclass[12pt,letterpaper]{article}

\usepackage[hang,perpage]{footmisc}

\usepackage{amsmath}
\usepackage{amsthm}
\usepackage{amssymb}
\usepackage{stmaryrd}
\usepackage{graphicx}
\usepackage{latexsym}
\usepackage{times}
\usepackage[varumlaut]{yfonts}
\usepackage[usenames]{color}
\usepackage{textcomp}
\usepackage{makeidx}
\usepackage{tocbibind}
\usepackage[bookmarks=true,backref=true,colorlinks=true,linkcolor=darkolivegreen,
citecolor=darkolivegreen,urlcolor=darkolivegreen]{hyperref}
\numberwithin{equation}{subsection}

\theoremstyle{definition}

\def\indexname{Index of terminology}
\makeindex

\newcommand{\captionfonts}{\footnotesize}
\makeatletter  
\long\def\@makecaption#1#2{%
  \vskip\abovecaptionskip
  \sbox\@tempboxa{{\captionfonts #1: #2}}%
  \ifdim \wd\@tempboxa >\hsize
    {\captionfonts #1: #2\par}
  \else
    \hbox to\hsize{\hfil\box\@tempboxa\hfil}%
  \fi
  \vskip\belowcaptionskip}
\makeatother   
\definecolor{darkolivegreen}{rgb}{0.333333, 0.419608, 0.1843140}

\makeatletter
\def\printnotation{{%
\def\indexname{Index of notation}
\begin{theindex}
\@input{\jobname.ntn}
\end{theindex}
}}
\makeatother

\makeglossary

\begin{document}

\title{{\bf Remarks on the relation between physics and faith}}

\author{Horst R. Beyer \\
 Louisiana State University (LSU) \\
 Center for Computation and Technology (CCT) \\
 330 Johnston Hall \\
 Baton Rouge, LA 70803, USA \\
 \& \\
 Max-Planck-Institute for Gravitational Physics \\
 Albert Einstein Institute \\
 Am M\"{u}hlenberg 1 \\
 D-14476 Golm, GERMANY}

\date{}                                     

\maketitle

\begin{abstract}
It is a quite common view among people, that
are not aware of the developments in modern physics, 
that it is part of human nature to substitute 
religious faith in places where there is no knowledge. 
Therefore, an increase in knowledge would lead to a decrease in 
the necessity of faith. Further, 
it is argued that, ideally speaking, a full knowledge of the 
laws of nature would make obsolete any sort of religious 
faith and would ultimately allow a complete control of nature 
by man. Since referring to nature, such views must be founded 
in the natural sciences of which physics 
is the most fundamental. Therefore, the question whether 
such views are compatible with the current state of 
natural sciences
is ultimately decided in physics. Indeed, it is likely that
these simplistic views have their origin in the world view 
generated by the successes of Newtonian physics from the middle
of the $17$th century until the beginning of the $20$th 
century which viewed the physical world as a type of mechanical 
clock in which the motion of the gears affect each other in a 
precise and predictable way. In particular,
the paper points out that the above 
views are no longer supported by current physics and
that abstracted world views cannot be 
considered as part of natural sciences, but only as belief 
systems.

\end{abstract}

\section{Introduction}

It is a quite common view among people, that
are not aware of the developments in modern physics, 
that it is part of human nature to substitute  
religious faith in places where there is no knowledge. 
Therefore, an increase in knowledge would lead to a decrease in 
the necessity of faith. A classical example is 
the medieval belief 
that the sun, moon, and stars were moved in their orbits by souls
(`anima'), 
who had to carry on this work until the last day. This belief was 
made obsolete by Kepler's laws of planetary motion. Further, 
it is argued that, ideally speaking, a full knowledge of the 
laws of nature would make obsolete any sort of religious 
faith and would ultimately allow a complete control of nature 
by man.

Since referring to nature, such views must be founded 
in the natural sciences of which physics 
is the most fundamental. Therefore, the question whether 
such views are compatible with the current state of 
natural sciences
is ultimately decided in physics. Indeed, it is likely that
the above simplistic views have their origin in the classical world view generated 
by the successes of Newtonian physics from the middle
of the $17$th century until the beginning of the $20$th 
century which viewed the physical world as a type of mechanical 
clock 
in which the motion of the gears affect each other in a precise 
and predictable way. It was assumed that with some more effort 
all the connections between those gears would soon be found. It 
is generally accepted that most physicists in the $19$th century 
believed that all the fundamental laws of nature were already 
known. In this world view there is no space for  
a Christian God or religious faith.

The paper points out that the above views are no longer 
supported by current physics  and
that abstracted world views cannot be 
considered as part of natural sciences, but only as belief 
systems. For this, it sketches the medieval world view as 
the starting point of the development of 
physics in the $17$th century,  
describes and analyzes that development and states 
cornerstones of Newtonian physics which  
have been assumed evident, but 
nevertheless have 
been destroyed by the Theory of Special 
Relativity, the Theory of General Relativity, Quantum Mechanics, 
Quantum Electrodynamics and the Standard Model of Particle Physics.

\section{The medieval world view}

At the beginning of the development of modern physics in 
the $17$th century by Johannes Kepler, 
Galileo Galilei and Issac Newton, 
the medieval world picture 
derived from the Ptolemaic universe
prevailed in Europe \cite{lewis}.

It consisted of a series of $10$ concentric transparent 
spheres or `heavens' with the earth at their center. Fixed 
to the first
$7$ innermost spheres were the `planets', i.e.,  
the moon, Mercury, Venus, the Sun, Mars, Jupiter and Saturn.   
Fixed to the $8$th sphere \footnote{The `Stellatum'.} were 
the stars. The $9$th sphere was called the `First Movable' 
\footnote{The `Primum Mobile'.} that carries no luminous body 
and therefore
is invisible. Finally, the $10$th sphere is `the very Heaven'
\footnote{The `Caelum ipsum'.} and full of the God, 
whose is considered immaterial and unaffected by time.
Hence God is unmovable, but the love for Him puts the 
the `First Movable' into rotation from east to west
and transfers its to the other spheres. 
The lower spheres are rotating more slowly from 
west to east and are forced back by the daily impulse
of the `First Movable'. All spheres 
are inhabited by `intelligences'. The terrestrial domain
of the universe inside the first innermost 
sphere contains bodies made out of the elements fire, air, 
water and earth. The heavenly bodies above the moon 
are made of a more perfect fifth element, the 
`Quintessence' or `aether'.

\section{The development of physics in 
the $17$th century}
 
\subsection{Description of the development}

In the following, we have a brief look at some of the 
scientific achievements 
of Kepler, Galilei, Newton and their influences on the philosophic 
and religious views in the last half of the $17$th century.

\begin{figure}
\centering
\includegraphics[width=3cm,height=4.1cm]{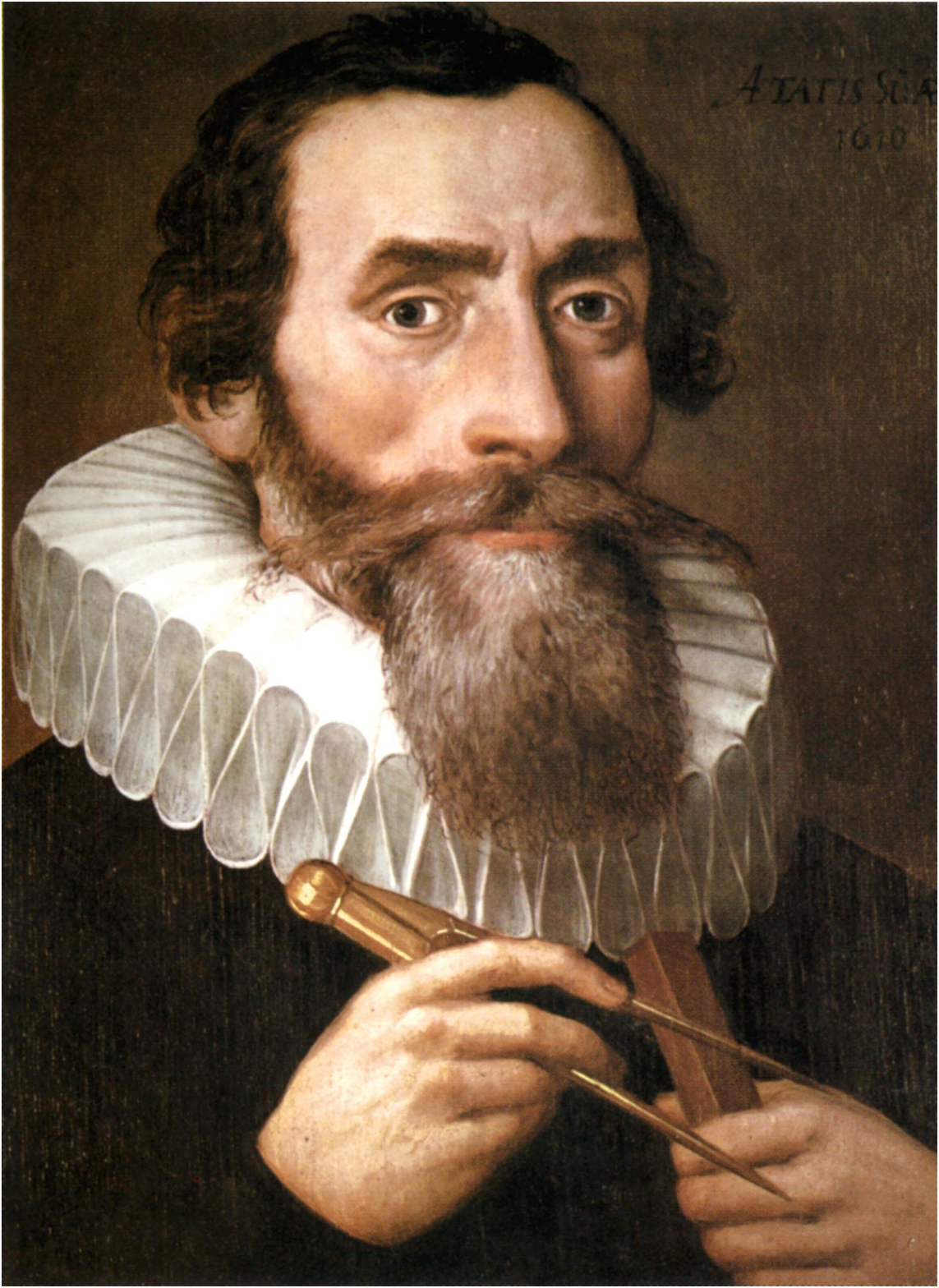}
\quad
\includegraphics[width=3cm,height=3.81cm]{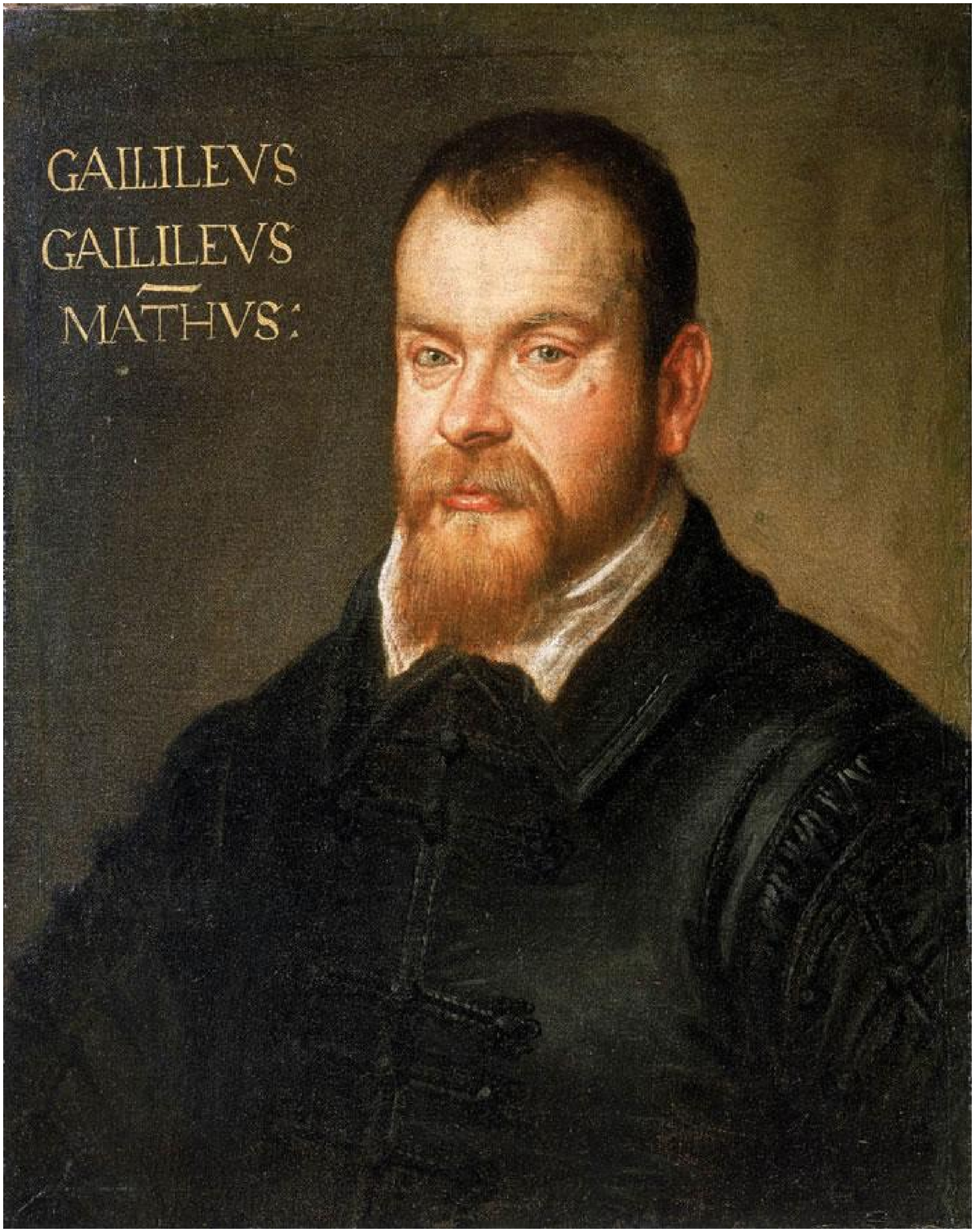}
\quad 
\includegraphics[width=3cm,height=4.12cm]{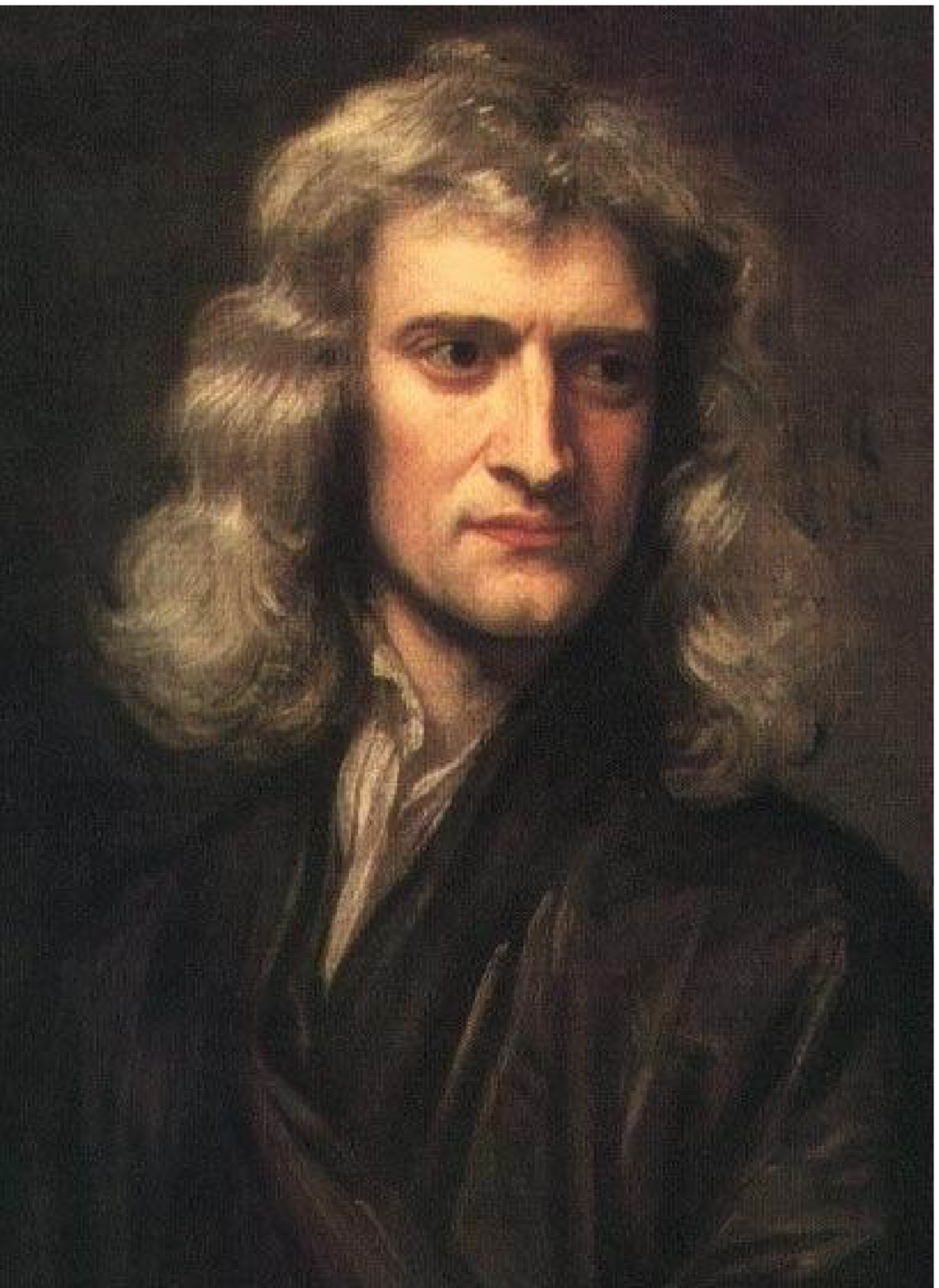}
\caption{Portraits of Kepler (1610), Galilei (ca. 1605-07) and 
Newton (1689).}
\label{fig1}
\end{figure}

{\it Kepler} $(1571-1630)$ was a devout Christian. 
One of Kepler's favorite biblical passages is

\begin{itemize}

\item 

John 1:14 (KJV)

\begin{quote}
``And the Word was made flesh, and dwelt among us, (and we beheld his glory, the glory as of the only begotten of the Father,) full of grace and truth.''
\end{quote}

\end{itemize}

For him, this indicates that the divine archetypes were 
made visible as geometric forms in nature. In particular, a 
sphere was a reflection of Christian Trinity with God the Father 
as its center, Christ the Son as periphery and the intervening 
space as Holy Spirit. He tried to explain the distances among 
the six Copernican planets \footnote{Mercury, Venus, Mars, Saturn, 
Jupiter and Uranus.} by circumscribing and 
inscribing each orbit with one of the five regular polyhedrons.\footnote{
The pyramid, cube, octahedron, dodecahedron, and icosahedron.}
In a letter from $1605$, he views the universe no longer as a divinely 
animated being, but as a clockwork 
\cite{kepler}
\begin{quote}
``Scopus meus hic est, ut Caelestem machinam dicam non esse instar divinj animalis,
sed instar horologij ($\cdot$ qui horologium credit esse animatum, is gloriam artificis tribuit operj $\cdot$), \dots \, .''
\end{quote}

One of Kepler's main 
scientific achievements are his three laws of planetary motion 
$(1609-1618)$. They 
are as follows: 
\begin{enumerate}
\item All planets move around the Sun 
in {\it elliptical} orbits, with the Sun 
in one of the foci. 
\item A radius vector joining a planet with the Sun sweeps out 
equal areas in equal lengths of time. 
\item The square of the time required by a planet
to complete one orbit  
is directly proportional to the cube 
of its mean distance from the Sun.    
\end{enumerate}

{\it Galilei} $(1564-1642)$ can be considered as the father 
of experimental sciences by performing precise measurements 
on mechanical systems with a subsequent mathematical analysis of 
the results. In particular, he studied balls rolling 
on inclined planes. From the results, he concluded that bodies do not need a proximate cause to stay in motion which  
contradicts the Aristotelian view that  
any motion presupposes a mover.
This result led, after 
generalization by Rene Descartes $(1596-1650)$ to motion on 
a straight line, 
to Newton's law of inertia. Timing the rate of the 
descent of balls, 
he deduced that freely falling bodies \footnote{in particular, not 
subject to the resistance of air} would be uniformly accelerated 
at a rate independent of their mass. Moreover, he understood 
that the 
motion of any projectile was the result of simultaneous and 
independent motion in the horizontal direction and falling 
motion in the vertical direction. In $1638$, Galilei wrote \cite{galileo2},
\begin{quote}
``It has been observed that missiles and projectiles describe a curved path of some sort; however no one has pointed out the fact that this path is a parabola. But this and other facts, not few in number or less worth knowing, I have succeeded in proving; ...''
\end{quote}

Finally, in the `Assayer' from $1623$, 
Galilei insisted that `the book of the Universe' is 
written in the language of mathematics \cite{galileo1}

\begin{quote}
``Philosophy is written in this grand book, the universe, which stands continually open to our gaze. But the book cannot be understood unless one first learns to comprehend the language and read the letters in which it is composed. It is written in the language of mathematics, and its characters are triangles, circles and other geometric figures without which it is humanly impossible to understand a single 
word of it; without these, one wanders about in a dark labyrinth.'' 
\end{quote}

Influenced by the results of Kepler, Galilei and later 
Newton, the dominant philosophy of the last half of the $17$th 
century was that of Rene Descartes. Mechanics is the basis of his physiology 
and medicine. {\it Descartes} believed that all material bodies, including the 
human body, are machines that operate by mechanical principles.
It is known that Newton studied the works of Descartes and other 
{\it mechanical philosophers}, who {\it viewed physical reality 
as composed entirely 
of particles of matter in motion and} who {\it held that all the 
phenomena of nature result from their mechanical interaction}.

In $1684$, {\it Newton} $(1643-1727)$ 
published his famous three laws of motion which
provide the basis of mechanics to this day \cite{newton}.  
Together with his and independently Gottfried Wilhelm Leibniz's 
$(1646-1716)$
discovery of differential and integral calculus, these 
laws transformed the field of mechanics into an exact science. 
In $1687$, Newton formulated his law of gravitation 
that have Kepler's laws of planetary 
motion and the results of Galilei's experiments on motion on 
inclined planes as particular consequences. Hence the laws of 
mechanics also unified the knowledge on mechanical systems at 
that time.    
    
\begin{figure}
\centering
\includegraphics[width=3cm,height=3.66cm]{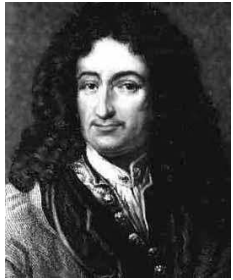}
\caption{Portrait of Leibniz, date unknown.}
\label{fig2}
\end{figure}

After the groundbreaking work of Newton, the motion of 
the planets and comets of the solar system was predictable to high
accuracy and understood on the basis of natural laws.
The announcement by Le Verrier in $1859$, that there was 
an unexplained advance in the perihelion of Mercury, was not 
regarded as a failure of Newtonian Mechanics, but due to unknown 
gravitational sources. An explanation for this advance was given 
by Einstein's Theory of General Relativity in $1916$ 
\cite{einstGR1}. A first indication for the breakdown 
of classical physics appeared at the end of $19$th century
in form of the failure of the Rayleigh-Jeans law 
to describe the observed behaviour of black body radiation. 
In $1900$, Max Planck  \footnote{In 
1918, Nobel prize winner in physics 
``in recognition of the services he rendered to the advancement of Physics by his discovery of energy quanta".}
derived his radiation law that gave a correct description, 
but a the cost of assuming that 
the oscillators comprising the blackbody could absorb 
energy only in discrete amounts, in quanta of energy. This 
led to
the development of Quantum Mechanics. 

\begin{figure}
\centering
\includegraphics[width=3cm,height=4.5cm]{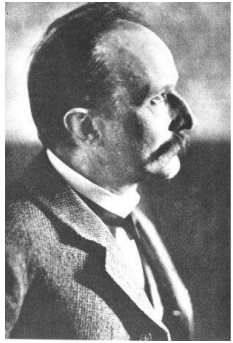}
\caption{Max Planck (1910).}
\label{fig3}
\end{figure}

{\it Because of its highly accurate descriptions, 
mechanics became regarded as the ultimate explanatory 
science. Phenomena of any kind, it was believed, could and 
should be explained in terms of mechanical concepts.
Newtonian physics was used to support the deistic view,
in particular represented by Voltaire $(1694-1778)$, 
that God had created the world as a perfect machine that then 
required no further interference from Him, the Newtonian world 
machine or Clockwork Universe.} 

The confidence in the validity 
of these ideas is best described  by the following statement 
\cite{laplace} of Laplace $(1749-1827)$ in $1814$:

\begin{figure}
\centering
\includegraphics[width=3cm,height=4.02cm]{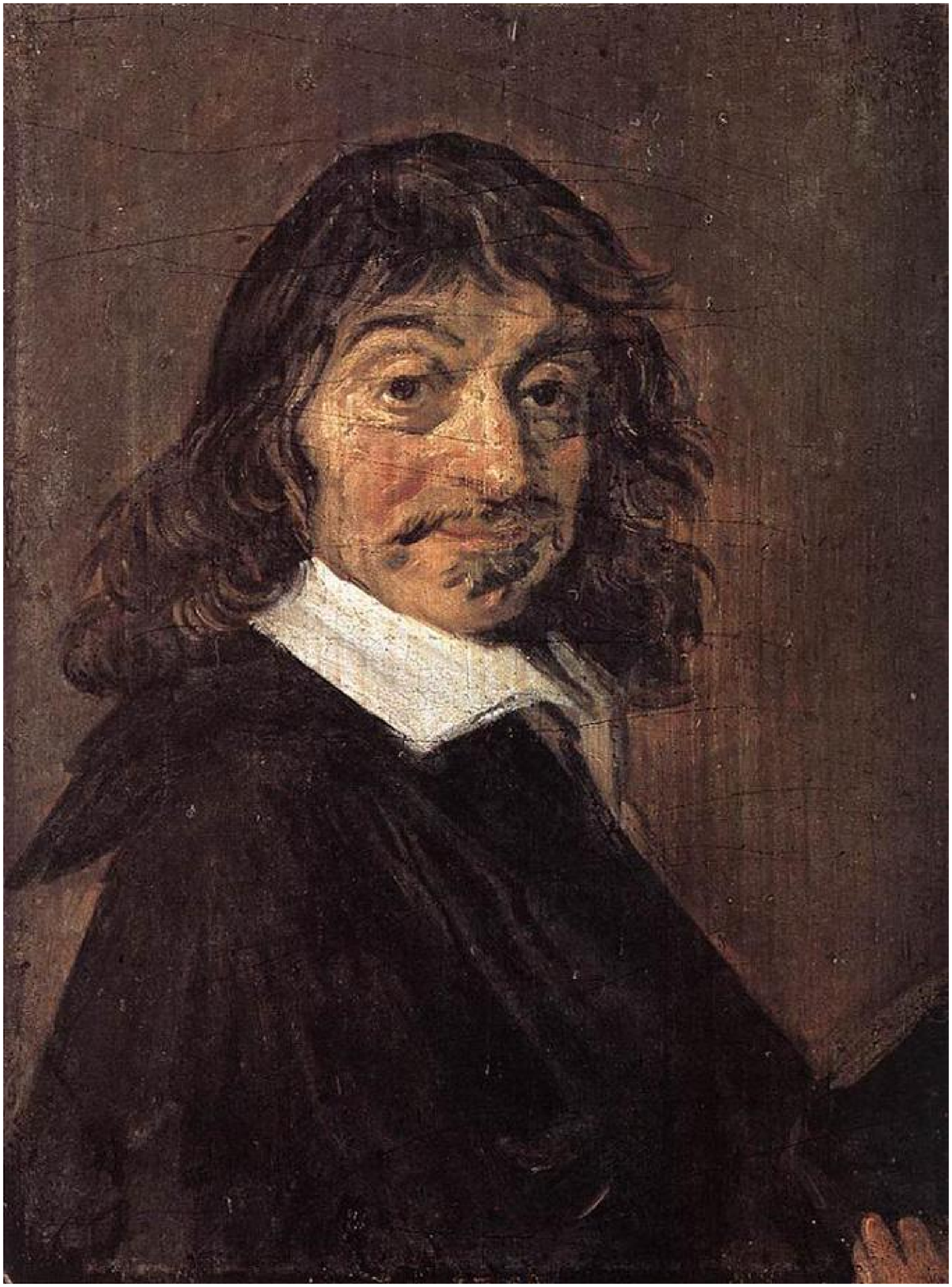}
\quad
\includegraphics[width=3cm,height=3.41cm]{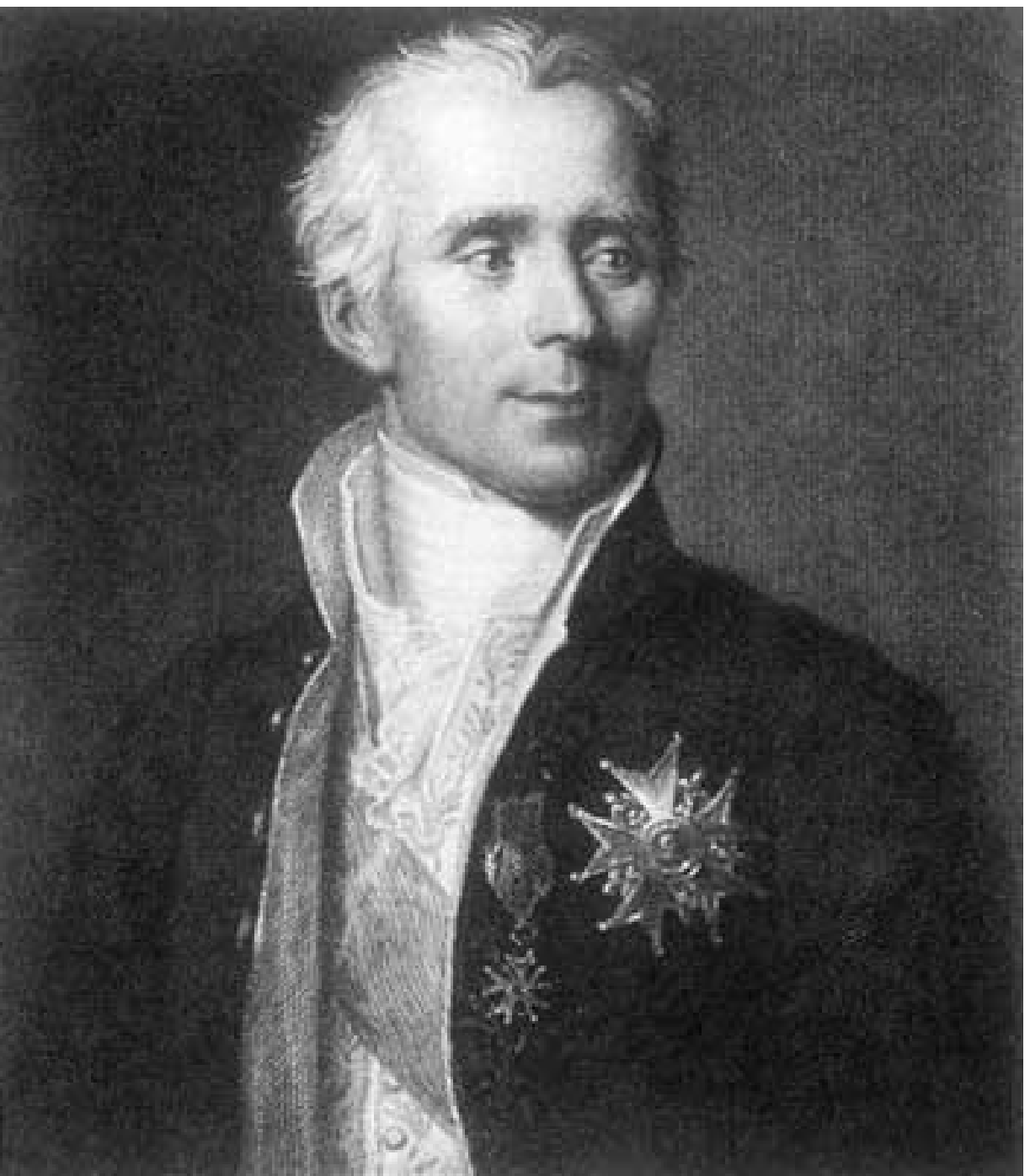}
\quad
\includegraphics[width=3cm,height=3.39cm]{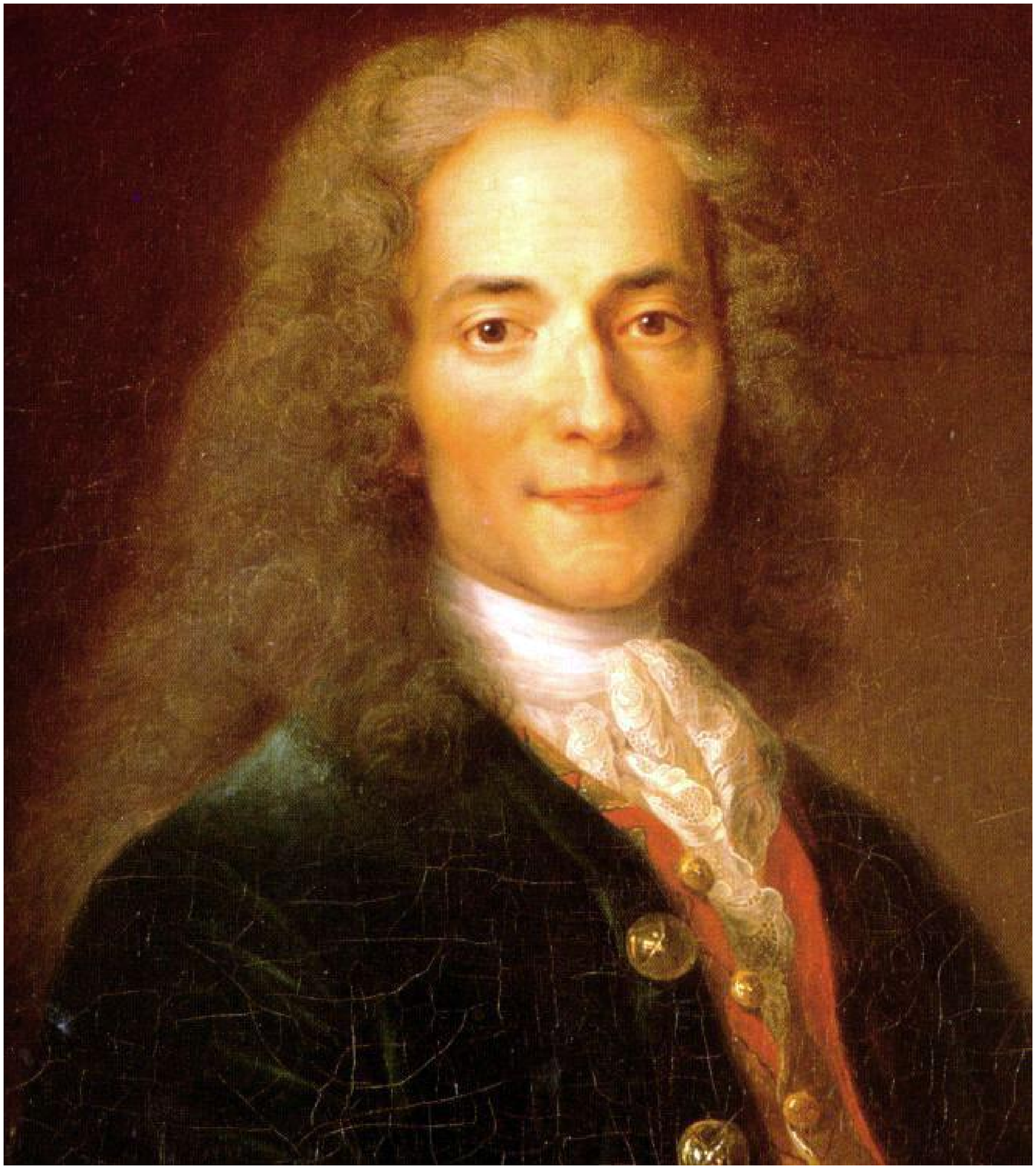}
\caption{From left to right: Portraits of Descartes 
(1649), Laplace and Voltaire (1718).}
\label{fig4}
\end{figure}

\begin{quote} 
``We ought then to regard the present state of the universe
as the effect of its anterior state and as the cause of the one 
which is to follow. Given for one instant an intelligence which 
could comprehend all the forces by which nature is animated 
and the respective situation of the beings who compose it --
an intelligence sufficiently vast to submit these data to 
analysis -- it would embrace in the same formula the movements
of the greatest bodies of the universe and those of the lightest 
atom; for it, nothing would be uncertain and the future, as the 
past, would be present to its eyes.''   
\end{quote}

In the aftermath, the method of Newtonian Mechanics was applied 
successfully to a growing number of areas of nature. The marvelous 
development of mechanics in the $18$th century, thermal engineering 
and thermodynamics in the beginning of the $19$th century give 
evidence of the power of that approach. 

\subsection{Analysis of the development}

In the following, from a modern perspective,
we analyze
the important changes in the approach to nature that have occurred 
in $17$th century. 

\subsubsection{The method of natural sciences}

First, the method of natural sciences had been established. This 
method tries to separate nature into systems that are to a good 
approximation `isolated', in the sense that the inclusion of their 
interaction with the remaining part of nature would introduce 
only small changes below the intended accuracy of description. 
Note that such separation is non-trivial and involves an 
{\em assumption} that itself cannot be verified. Such 
supposedly isolated 
systems are studied by precise experiments or 
observations. The corresponding measurements are subjected to a 
mathematical analysis with the goal of abstraction of a Natural 
Law or `Theory'. Predictions of the theory are tested by 
additional experiments or observations. If the results are 
consistent with those predictions within the 
accuracy of the measurements, the theory is 
maintained. Otherwise, the theory is revised.  
In the last case, the process is continued until consistency with 
known experiments and observations is reached. 

Note 
that even if the initial assumption of isolation of the system is 
false, this process can still lead to an accurate 
description by the theory if, for some reason, 
it successfully `captures' the influence
of the surrounding. The simplest way to 
achieve this
is to include a larger number of constants into the theory 
that can be adapted to measurements. To avoid the last, 
it is generally demanded that a theory should be in some
sense `simple' or `aesthetic'.

Experiments and observations provide the basis of the method 
of natural sciences in 
that their outcome decides on the `truth' of a theory.  
If predicted qualitatively and quantitatively correctly 
by a theory, 
a natural phenomenon is considered to be `understood' or 
`explained' {\it in terms of the theory}. 

Note that in the case 
of a false assumption of isolation of the system, a `true' theory 
is still `false' in the sense that it cannot be expected that 
it describes correctly the `nature' of the system. Note also that 
{\it there are no proofs in natural sciences}, although Galilei 
uses this notion in the above cited passage from $1638$.\footnote{This is not in conflict with the fact that  
theoretical physics describes physical systems by 
mathematical means.
In this way, physical problems are mapped 
onto mathematical 
problems. Inevitably, the last  
need to be solved by pure mathematical methods, 
i.e., by theorems and proofs.} 
Differently
to mathematics, those are based on experiment, not on logic. 
Ultimately, in natural sciences even logic is subject to 
experimental verification.

\subsubsection{Successes and abstracted world 
views}

The focus on mechanical phenomena allowed the development 
of a highly successful theory, i.e., Newtonian Mechanics, that was 
believed to be 
consistent with experiments and observations of increasing 
accuracy until the beginning of the $20$th century.\footnote{
The announcement by Le Verrier in $1859$ that there was 
an unexplained advance in the perihelion of Mercury was not 
regarded as a failure of Newtonian Mechanics, but due to unknown 
gravitational sources.}
Under the impression of the successes of this theory, there was abstracted 
a philosophical world view, the Clockwork Universe where God is 
regarded as a creator of natural laws governing its evolution and 
the agent that set the clock initially into 
motion. After that, the evolution of the world 
is thought to be completely deterministic. It is believed that 
a miracle, in the 
sense of a divine 
intervention into nature which does contradict natural 
laws, does not occur.

\subsubsection{Abstraction of 
world views from natural sciences}

That abstraction of a philosophical world view, although tempting, 
provides a classic example of the quite 
common {\it mistake} of the layman {\it to identify a theory with 
the part of nature it only describes}.
The method of natural sciences {\it defines} the `truth' of a  
theory as its consistency with known experiments and observations 
within the accuracy of measurements. There is no
generally accepted method available to decide whether the theory is 
true in some absolute sense. On the contrary, 
the history of physics showed that ultimately 
every theory turned out to be `false', in the sense that 
its predictions were inconsistent with more precise 
experiments or observations which made its revision 
necessary. Such revisions led to new 
theories which explained the phenomena of the older theories 
qualitatively and quantitatively and at the same a large number 
of additional natural phenomena those older theories could not 
account for. {\it But, as will be pointed out later on, 
in every step in this sequence, concepts that were fundamental 
in the formulation of the older theory were negated by its 
successor. From this point of view, it can very well be said 
that fundamental aspects 
of the older theories were false.} There is no reason to believe
that this will be different for future theories.
Here, it should also be taken into account that 
by the rules of logic, a false statement can imply  
a true statement. For instance, a compound statement falsely 
claiming a true statement and a false statement at the same
time leads to a true statement by omitting that false part 
of the statement. The statement 

\begin{quote}
{\it `Tigers are mammals and mice are reptiles'}
\end{quote} 

is false, but it implies the true statement 

\begin{quote}
{\it `Tigers are mammals'}.
\end{quote}

Hence, a false theory can very 
well lead to results consistent with experiments or observations.
For these 
reasons, the abstraction of philosophical a world view from such 
theories is inherently flawed.

{\it Therefore, the abstraction 
of philosophical world views from natural sciences leads to a belief 
system, only. In particular, differently from these sciences, 
such views cannot claim exactness.} 

In addition, from a Christian point of view, it appears that 
the creation of a philosophic world view, {\it by abstraction from
natural sciences or by theological reasoning from statements in  
the bible}, violates commandments in the bible:

\begin{itemize}     
      
\item 
Exodus 20 (KJV)
\begin{quote}
``4 Thou shalt not make unto thee any graven image, or any likeness of any thing that is in heaven above, or that is in the earth beneath, or that is in the water under the earth. 5 Thou shalt not bow down thyself to them, nor serve them: ...''
\end{quote}
\item 
or in Deuteronomy 5 (KJV)

\begin{quote}
``8 Thou shalt not make thee any graven image, or any likeness of any thing that is in heaven above, or that is in the earth beneath, or that is in the waters beneath the earth: 9 Thou shalt not bow down thyself unto them, nor serve them: ...''
\end{quote}

\end{itemize}

\subsection{A general remark on Darwin's 
theory of evolution}

Since I am no biologist, I will make only some very general remarks 
on the relation between physics and
the natural sciences of biology and chemistry. These start 
from definitions of these 
fields given in the Encyclopedia Britannica \cite{britannica}:

\begin{quote} 
``Biology is the study of living things and their vital 
processes.'' 
\end{quote}

A large part of those processes have been recognized as 
chemical processes which led to the creation 
and growing importance 
of the subfields of Biochemistry and Molecular Biology.
Judging from the successes of those, it does not seem to
be too far fetched to say that biology will ultimately turn 
out to be (at least in principle) a subdivision of chemistry. 

\begin{quote}
[Chemistry is] ``the science that deals with the properties, composition, 
and structure of substances (elements and compounds), the reactions and 
transformations they undergo, and the energy released or absorbed during 
those processes.''
\end{quote}

The laws that govern those properties of substances have been 
recognized as the laws of quantum mechanics. Hence, it can be said 
that chemistry is (at least in principle) a subdivision of physics. 

As a consequence, physics increasingly emerges as the root of the 
natural sciences.
Therefore, there is reason to view Darwin's and other theories of 
evolution
from a physicist's point of view. Presently, because of their 
ad hoc nature and 
lack of quantitative statements, from this point of view, those theories 
seem to be more working hypotheses than theories. Also for this reason, they seem 
to be unsuitable for the abstraction of a world view.

\section{The descent of the mechanical theory of nature}

In the following, we state the 
established physical theories subsequent to 
Newtonian Mechanics in chronological order and give 
a short description of fundamental changes they introduced.

\subsection{Maxwell's Theory of the electromagnetic field}

That descent of the mechanical theory of nature 
started with the analysis of electrical and magnetic
phenomena in the $19$th century, in particular with the discovery 
of radio waves propagating 
at the speed 
of light in vacuum.\footnote{i.e., at the speed of 
$299,792,458$ meters per second} 
Such waves were predicted by the electromagnetic 
theory of James Clerk Maxwell $(1831 - 1879)$ from $1873$ \cite{maxwell}. 
They were produced and measured in the laboratory by Heinrich Hertz 
$(1857 - 1894)$ in $1887$. 

\begin{figure}
\centering
\includegraphics[width=3cm,height=3.9cm]{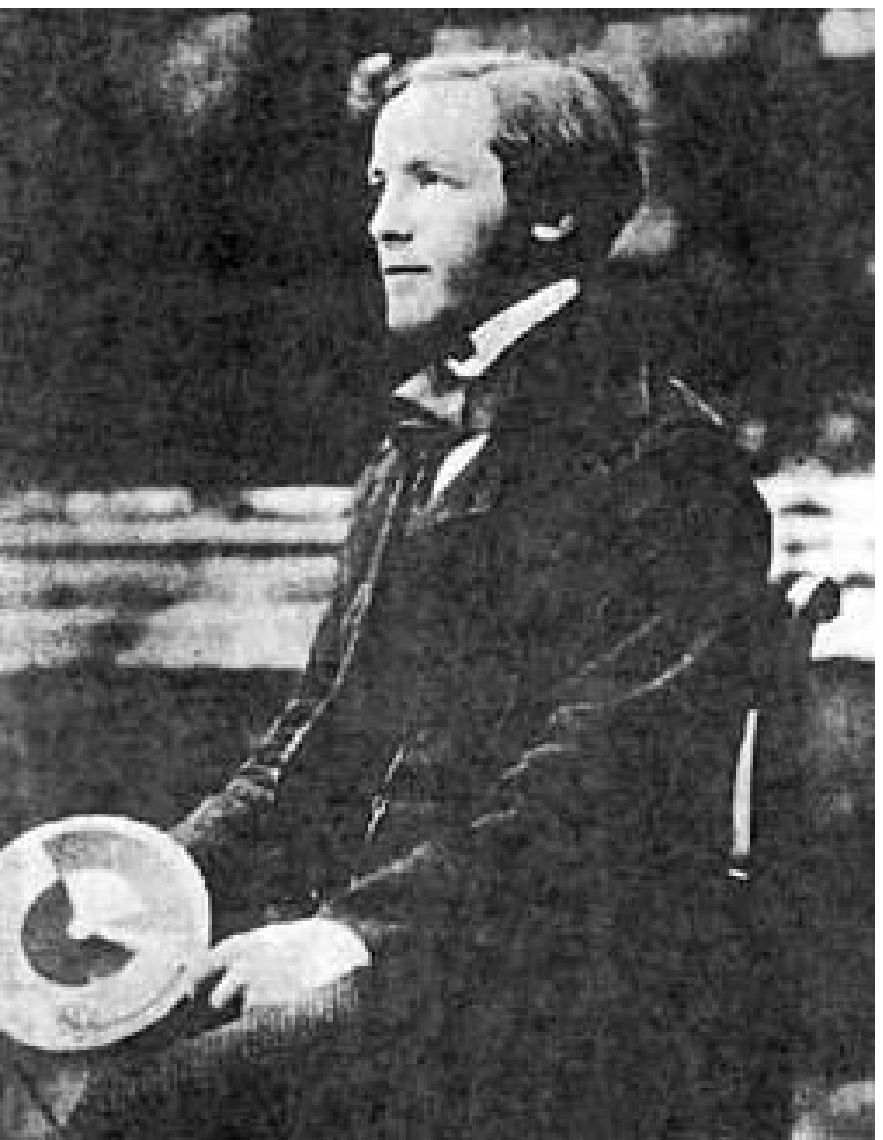}
\quad
\includegraphics[width=3cm,height=3.02cm]{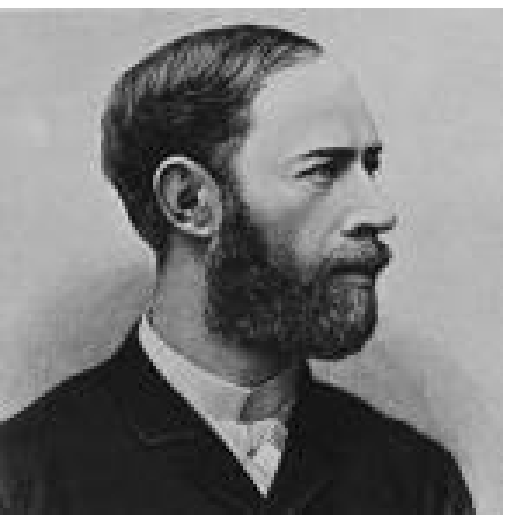}
\caption{From left to right: Maxwell and Hertz, dates unknown.}
\label{fig5}
\end{figure}

It is worth noting that in attempting to illustrate 
Faraday's law of induction \footnote{that a changing magnetic field 
gives rise to an induced electric field}, Maxwell constructed a mechanical 
model \cite{maxwell}. Also, it was assumed that those waves had some sort 
of a material carrier, the so called `ether', similar to sound waves that 
are compression waves in matter. In $1881$ and $1887$, this theory was 
shattered by the Michelson-Morley experiments, which were designed 
specifically to detect the motion of the Earth through the ether and which 
indicated that there was no such effect.

{\it As a consequence, the existence of non-mechanical entities in 
addition to 
matter was realized, i.e., force fields that are propagating through 
empty space.}

\subsection{The Special Theory of Relativity}
 
Maxwell's electromagnetic equations require waves to 
move at the speed of light in vacuum in every inertial system, that 
is in every reference frame where free \footnote{i.e., free from 
the action of external forces} matter moves with  
constant speeds and on a straight lines. Two such reference frames  
move relative to each other at a constant speed. Therefore, the 
propagation of electromagnetic waves contradicts 
common experience \footnote{of motion that 
is small compared to the speed of light in vacuum} 
that suggests that the wave speed should 
appear to be different as measured from different inertial systems.

\begin{figure}
\centering
\includegraphics[width=3cm,height=3.68cm]{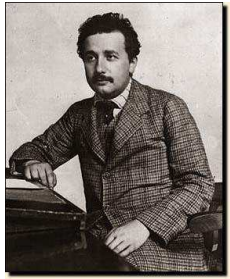}
\caption{Albert Einstein, ca. 1905.}
\label{fig6}
\end{figure}

In $1905$, Albert Einstein $(1879-1955)$ \footnote{In 
1921, Nobel prize winner in physics ``for his services to Theoretical Physics, and especially for his discovery of the law of the photoelectric effect".}
made the constancy of the speed 
of light and the requirement that the laws of physics should assume 
the same form 
in every inertial system {\it postulates} of his Special Theory of 
Relativity \cite{einstSR}. This led to radical changes of 
Newtonian Mechanics. The concept of absolute time or rather 
of absolute simultaneity of events lost its meaning 
and became dependent on the reference frame instead. As 
a consequence, also the concept of rigid bodies had to be 
abandoned.  
{\it All these notions are 
central in Newtonian Mechanics}.

\subsection{The Theory of General Relativity}
 
Newton's law of gravitation is incompatible with the Theory of 
Special Relativity and therefore needed a revision after 
the arrival of this theory. This was achieved in $1915$ by Albert 
Einstein's Theory of General Relativity \cite{einstGR,einstGR1}
that absorbs the 
gravitational field into the geometry of space-time. In this, 
the theory generalizes the Special Theory of Relativity
and again radically alters Newtonian Mechanics on a fundamental 
level. In particular, it removes the gravitational field from any 
energy momentum balance of a physical system under the influence 
of such a `field'. As a consequence, in general there is no
conserved total energy or momentum associated to such a system.

\subsection{Quantum Mechanics}

\begin{figure}
\centering
\includegraphics[width=3cm,height=4.05cm]{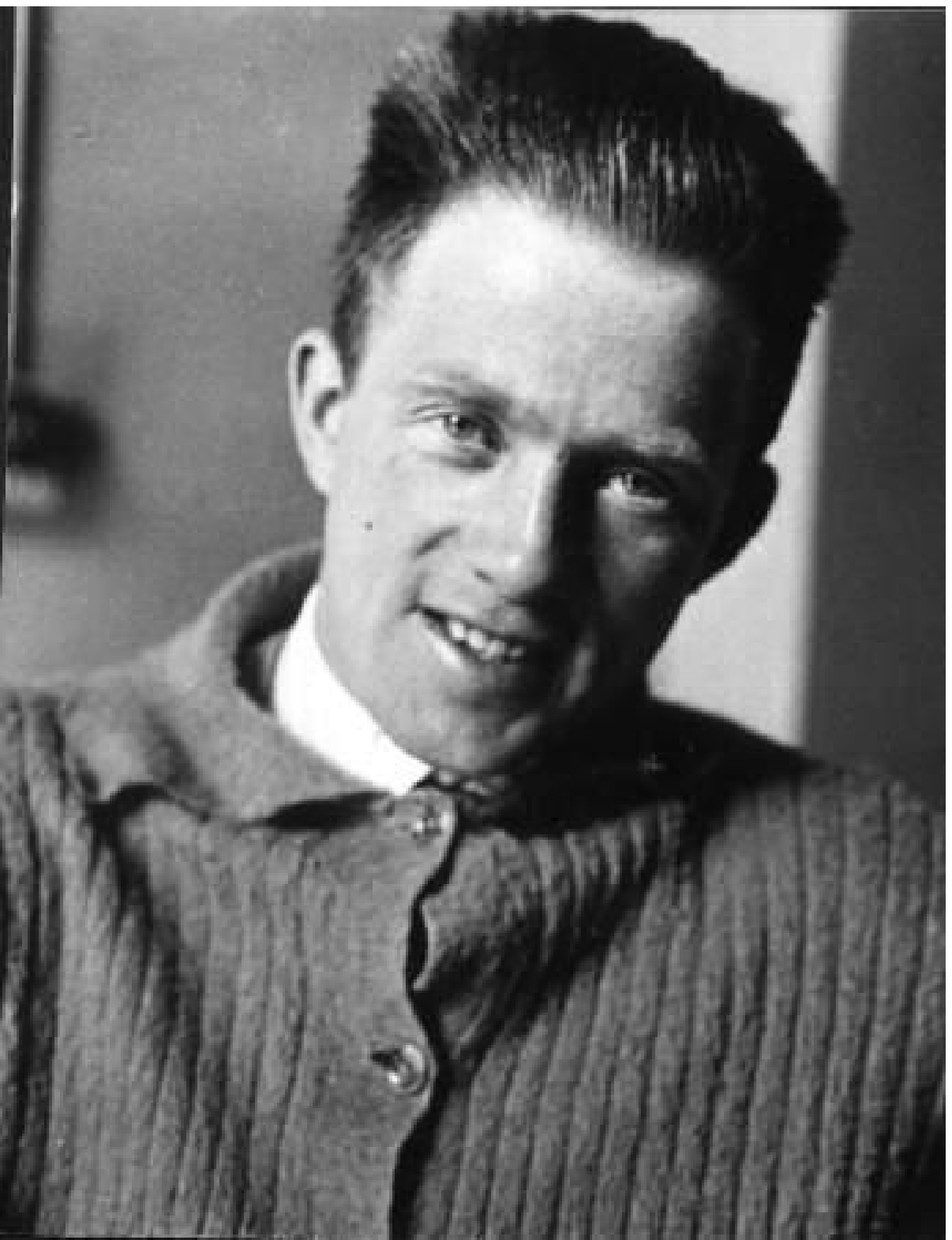}
\quad
\includegraphics[width=3cm,height=4.17cm]{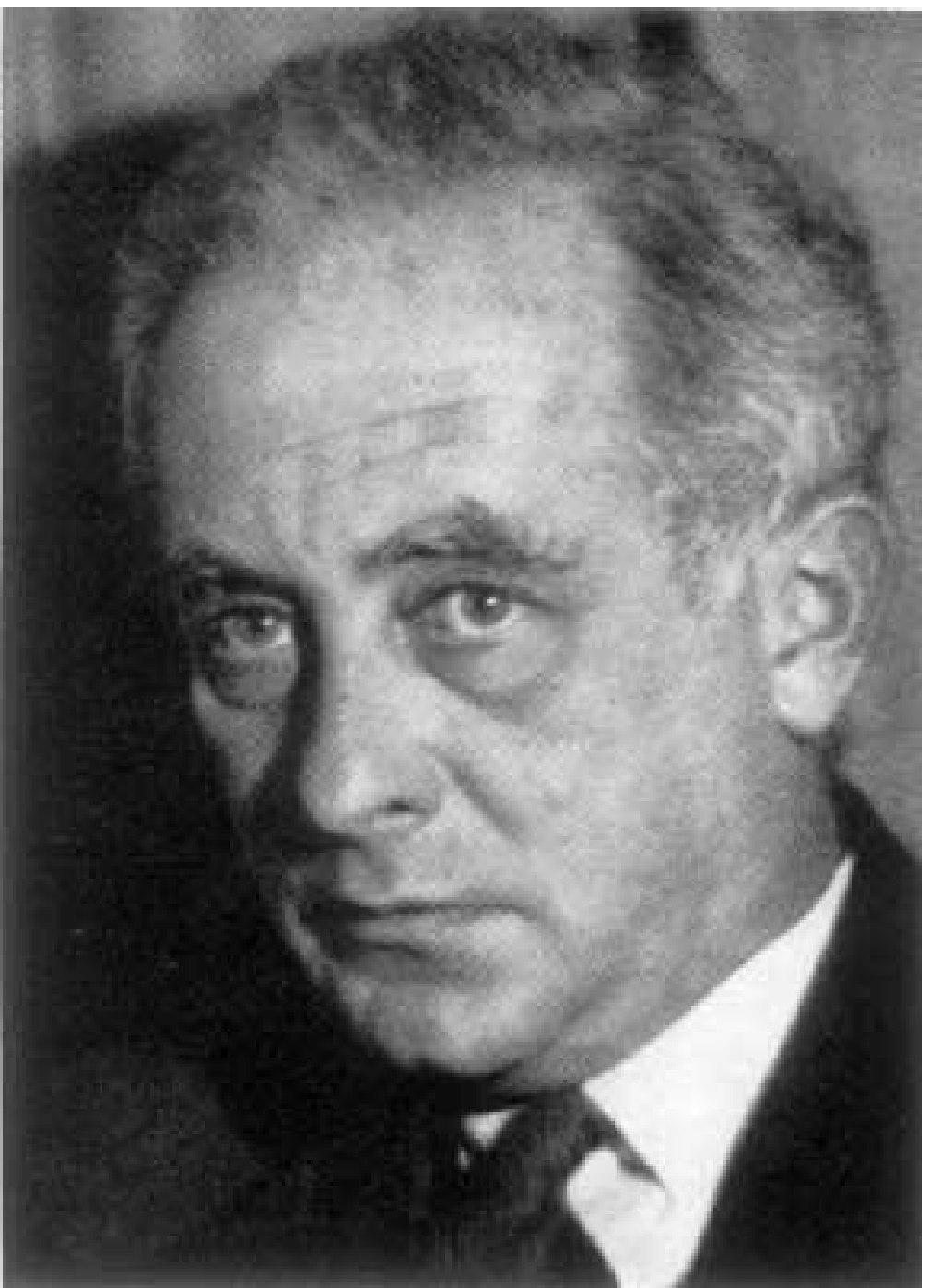}
\quad 
\includegraphics[width=3cm,height=3.82cm]{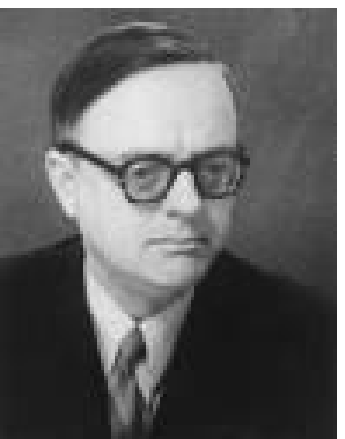}
\caption{From left to right: Heisenberg (ca. 1924), Born and 
Jordan.}
\label{fig7}
\end{figure} 

The Theory of Quantum Mechanics or more general Quantum Theory
was developed by Werner Heisenberg $(1901-1976)$
\footnote{In 1932, Nobel prize winner in physics 
``for the creation of quantum mechanics, the application of which 
has, inter alia, led to the discovery of the allotropic forms of hydrogen".}, Max Born 
$(1882-1970)$ \footnote{In 1954, shared with Walther Bothe, 
Nobel prize winner in physics
``for his fundamental research in quantum mechanics, especially for his statistical interpretation of the wavefunction".}
and Pascual Jordan $(1902-1980)$ in $1925$ because 
of the clear failure of Newtonian Mechanics to describe the 
physical phenomena in the range of atomic distances, i.e, 
$10^{-10}$ meter $= 0.1$ nanometer and below \cite{heisenberg1, 
born,heisenberg2}. This theory shook 
the foundations of physics like no other before or after 
because of its probabilistic character. That 
character allows statements on 
individual systems only in the exceptional cases of the 
occurrence of probabilities of value $1$. 

{\it It denies the possibility of a deterministic 
description  simultaneously of the positions and the 
momenta of the members 
of a mechanical system. It gives a  
deterministic description only for the probability distribution
of those quantities.} 

In this, it departs radically from the previous theories 
which were from then on referred to as 
`classical'.  The statistical nature of the 
theory was a major shock for physicists and is the reason why 
a few physicists still consider it to be an incomplete theory.
Notably, Albert Einstein was among them. 

The desperation of physicists, caused by the inapplicability
of the ideas from classical physics to atomic systems, 
is very well depicted in a letter from 
$1925$ of 
Wolfgang Pauli $1900-1958$,  
later a Nobel Prize winner in physics \footnote{in 1945, 
"for the discovery of the Exclusion Principle, also called the Pauli Principle".}, to his 
assistant Ralph Kronig $1904-1995$: 

\begin{quote}
`Currently, Physics is again very muddled. For me 
at least, it is to complicated and I wished I would be a 
comic artist of the screen or similar and would never have 
heard anything from physics. But still I hope that Bohr will 
save us with a new idea.' 
\end{quote}

\begin{figure}
\centering
\includegraphics[width=3cm,height=4.24cm]{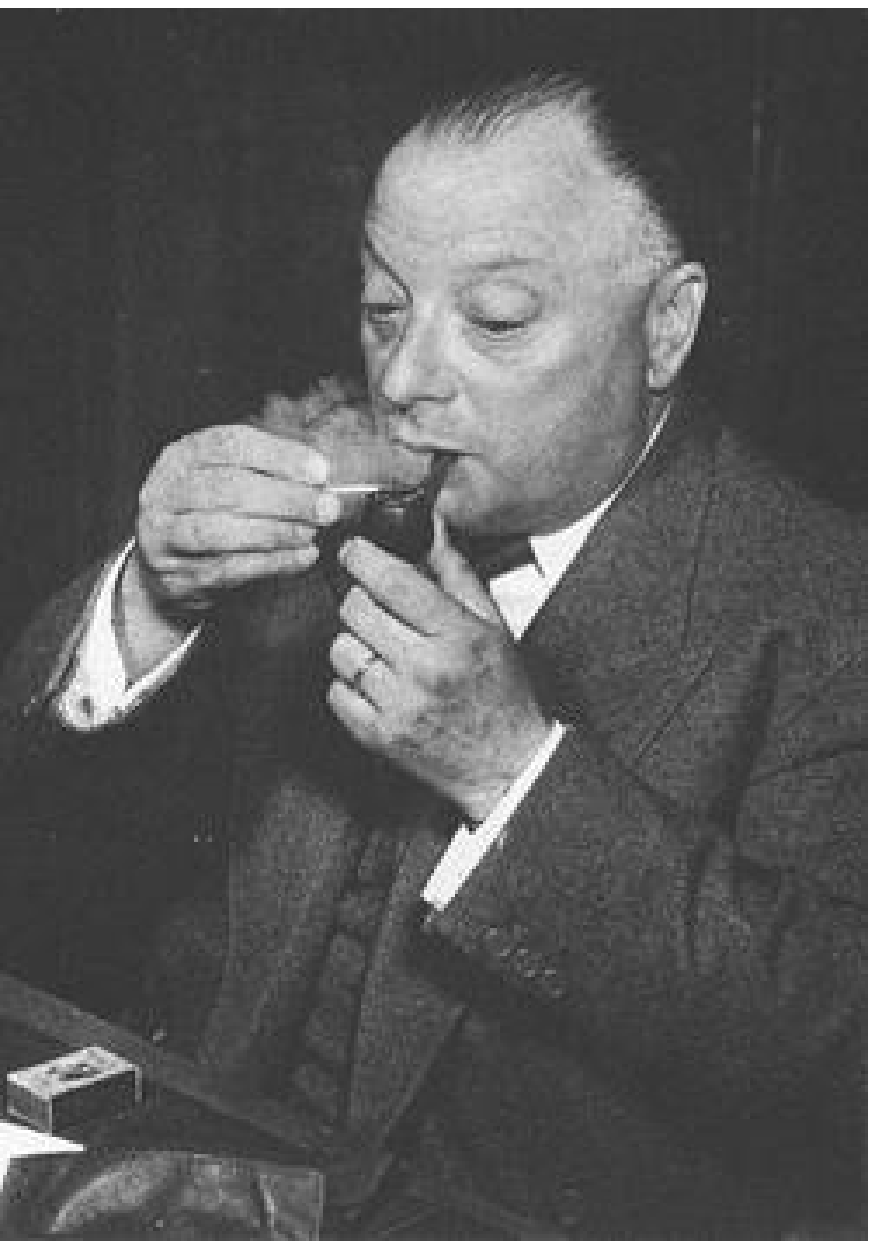}
\quad
\includegraphics[width=3cm,height=4.47cm]{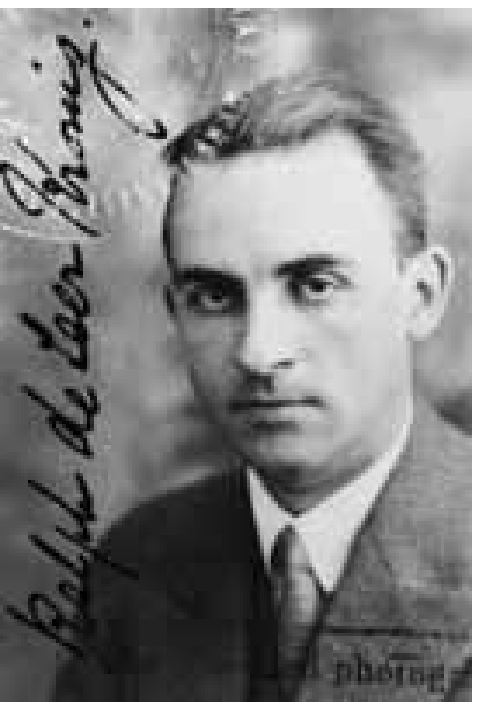}
\caption{From left to right: Pauli and Kronig.}
\label{fig8}
\end{figure}

\subsection{Quantum Electrodynamics}
Quantum Electrodynamics is the result of the attempt to develop 
a Quantum Mechanics which is compatible with the Theory 
of Special Relativity. It describes the interaction of charged 
matter. In this theory, 
matter and the electrodynamic field are treated on the same 
footing, 
namely as objects that can act as particles or fields. 
{\it In classical 
physics both properties are incompatible}.
In addition, it leads on the existence of antimatter that 
in collision with the corresponding matter annihilates to 
light. Also the opposite process is possible. As a consequence, 
matter and light can be considered as two manifestations of 
one field.
Some of the most precise tests of Quantum Electrodynamics
have been experiments 
dealing with the properties of subatomic particles known as muons. 
The magnetic moment of this type of particle has been shown to 
agree with the theory to nine significant digits. Agreement of 
such high accuracy makes Quantum Electrodynamics one of the most 
successful physical theories so far devised.

\subsection{The Standard Model of Particle Physics}

The Standard Model of Particle Physics is a generalization of 
Quantum Electrodynamics which also includes weak and strong 
nuclear forces that are in particular responsible for radioactive 
decay and the stability of atomic nuclei. It does not 
take into account the gravitational field. Until 
recently, it has proved highly successful
in the description of experimental observations.  
However, experiments in $1998$, (K2K-I) $2004$ (K2K-II) 
and $2006$ (MINOS) indicate that 
neutrinos have mass which is not taken into account 
in the standard model. 

Also, one of the basic assumptions of particle physics in general 
is the ancient idea (Democritus, ca. 
$460-370$ BC) of the existence of `atoms' \footnote{The Greek
word `atomiki' means `indivisibles'.}, i.e., of 
particles that are `fundamental' in the 
sense of not being composites of others. 
The history of particle 
physics does not seem to support this idea.

\section{Conclusions}

I hope, it has been made clear that natural sciences 
pose {\it no threat to faith because of the inherent limitations
of their approach}. Abstracted world views 
pose a potential threat, but leave the firm ground of
natural sciences and therefore constitute belief systems
that cannot claim exactness. 

That such abstractions are inherently flawed can also 
be read off from the development 
of the root of natural sciences, i.e., physics. Its development
led to a sequence of theories which give the impression of 
insights into nature of increasing deepness and comprehensiveness.
Still, each new theory negated fundamental aspects of its 
predecessor. It is not to be expected that this will change 
in future. 
This indicates that every physical theory contains some 
fundamentally false aspects which would be inherited by an 
abstracted world view. 

In addition, it should also be taken into account that the
majority of the creators of modern physics, 
in particular, 

\begin{quote}
Kepler, Galilei, Newton, Descartes, Laplace, 
Leibniz, Maxwell, Planck, Heisenberg and Jordan
\end{quote}

were Christians that did not see any contradiction between their 
work and their faith. Finally, I see no better way to conclude 
this paper with the final 
words of Max Planck in his speech ``Religion und Naturwissenschaft''
\cite{planck} (\,Religion and Natural Sciences) from 1937:

\begin{center}
{\bf 
Hin zu Gott!}
\end{center}

which translates to 

\begin{center}
{\bf Towards God!}
\end{center}


\begin{thebibliography}{99}


\bibitem{born} 
Born M, Jordan P 1925, {\em Zur Quantenmechanik}, 
Zeitschrift f\"{u}r Physik, {\bf 34}, 858-888.


\bibitem{britannica} 
Britannica Concise Encyclopedia 2006.

\bibitem{heisenberg2}
Born M, Heisenberg W, Jordan P 1925,
{\em Zur Quantenmechanik II}, Zeitschrift f\"{u}r Physik, 
{\bf 35}, 557-615.

\bibitem{borz} Borzeszkowski H-H, Wahsner R 1980,
{\em Newton und Voltaire}, Berlin: Akademie Verlag.

\bibitem{chojecka} Chojecka E 1967, {\em Johann Kepler und 
die Kunst. Zum Verh\"{a}ltnis von Kunst und Naturwissenschaften in 
der Sp\"{a}trenaissance}, Zeitschrift f\"{u}r Kunstgeschichte, 
{\bf 30}, 55-72.

\bibitem{cohen} Cohen I B 1983, {\em 
The Newtonian revolution}, Cambridge: University Press.

\bibitem{copernicus} Copernicus N 1995, {\em On the revolutions of
heavenly spheres}, New York: Prometheus Books.

\bibitem{dijksterhuis} Dijksterhuis E J 1985, 
{\em The mechanization 
of the world picture}, Princeton: Princeton University Press.
 
\bibitem{einstSR} Einstein A 1905, {\em Zur Elekrodynamik 
bewegter K\"{o}rper}, Annalen der Physik, {\bf 17}, 891-921.

\bibitem{einstGR}
Einstein A 1915, {\em Die Feldgleichungen der Gravitation},
Sitzungsberichte der 
der Preussischen Akademie der Wissenschaften 
zu Berlin, 844-847.

\bibitem{einstGR1} Einstein A 1916, 
{\em Die Grundlage der allgemeinen Relativit\"{a}tstheorie}, 
Annalen der Physik, {\bf 49}, 769-822.

\bibitem{einst} Einstein A 2005, {\em Mein Weltbild}, 
Z\"{u}rich: Europa Verlag.

\bibitem{galileo1} Galilei G, 
{\em Discoveries and Opinions of Galileo}, 
Translated with an introduction and notes by 
Stillman Drake, New York: Anchor Books. 

\bibitem{galileo2} Galileo G 2002, 
{\em Dialogues concerning two new sciences}, Philadelphia: Running Press.

\bibitem{heisenberg} Heisenberg W 1971, {\em Schritte 
\"{u}ber Grenzen}, M\"{u}nchen: Piper \& Co. Verlag.

\bibitem{heisenberg1}
Heisenberg W 1925, {\em 
\"{U}ber quantentheoretische Umdeutung kinematischer und 
mechanischer Beziehungen}, Zeitschrift f\"{u}r Physik, 
{\bf 33}, 879-893. 

\bibitem{jordan} Jordan P 1963, {\em Der Naturwissenschaftler 
vor der religi\"{o}sen Frage}, Oldenburg: Gerald Stalling Verlag.

\bibitem{kepler} Kepler J, Dyck W 1951, {\em Gesammelte Werke}, 
Vol. 15, p. 146. In a letter to Herwart von Hohenburg from 10th of April 1605.

\bibitem{kepler1} Kepler J 2002, {\em Harmonies of the world},
London: Running Press.

\bibitem{koestler} Koestler A 1968, {\em The sleepwalkers}, 
London: Penguin Books.

\bibitem{laplace}  Laplace P S 1995,
{\em A philosophical essay on probabilities}, 
New York: Dover Publications.

\bibitem{lewis} Lewis C S 1994, {\em The Discarded Image: An Introduction to 
Medieval and Renaissance Literature}, Reprint edition,  
Cambridge: Cambridge University Press.

\bibitem{newton} Newton I 1999, {\em The Principia}, 
Berkeley: University of California Press.

\bibitem{cosmosbiostheos} Margenau H, Varghese R A 1992, 
{\em Cosmos, Bios, Theos}, Chicago: Open Court Publishing. 

\bibitem{maxwell}
Maxwell J C 1891, {\em A Treatise on Electricity and Magnetism},
London: Clarendon Press.

\bibitem{planck} Planck M 1937, {\em Religion und Naturwissenschaft}, in: Planck M, {\em Wege zur physikalischen Erkenntnis}, 
4th ed., Leipzig: Hirzel Verlag.

\bibitem{pauli}
Enz C P, v. Meyenn K (eds.), 
{\em Wolfgang Pauli - Wissenschaftlicher Briefwechsel},
Vol. 1, Letter 89, 215.

\bibitem{strong} Strong E W 1952, {\em Newton and God}, 
Journal of the History of Ideas, {\bf 13}, 147-167.

\bibitem{voltaire} Voltaire 1978, {\em Letters on England},
London: Penguin Books. 

\end{thebibliography}
\end{document}